# On the Properties of Elemental and High-$T_c$ Superconductors in an Applied Magnetic Field in a Unified Framework

G.P. Malik[1#*] V.S. Varma[2#]

[1]*School of Environmental Sciences, Jawaharlal Nehru University, New Delhi 110067, India*

[2]*Department of Physics and Astrophysics, University of Delhi, Delhi 110007, India*

[1]*E-mail:* gulshanpmalik@yahoo.com

[2]*E-mail:* varma2@gmail.com

### *Abstract*

*The unified approach based on the Generalized BCS equations incorporating chemical potential (μ) employed to deal with the critical temperature, gap(s) and coherence length(s) of any superconductor (SC) in an earlier paper is shown here to be also applicable when the SC is in an applied field. Presented herein are the calculated values of the following parameters related to its penetration depth and critical current density: the interaction parameter governing the formation of Cooper pairs (CPs), the number of occupied Landau levels, the number density of charge carriers, the critical velocity of CPs, and the values of μ at $T \approx T_c$ and zero. Our study is found to corroborate the finding reported by Rasolt and Tešanović [Revs. Mod. Phys., 64, 709 (1992)] that in some systems the effective electron-electron interaction is enhanced with increasing magnetic field and sheds new light on the finding reported by Audouard et al. [Euro. Phys. Lett., 109, 27003 (2015)] that the properties of a superconductor in magnetic fields are controlled by a single band despite the multiband nature of the Fermi surface. The SCs dealt with are Cd, Zn, Al, In, Hg, $MgB_2$, YBCO, Bi-2212 Bi-2223, Tl-2212, Tl-2223 and compressed $H_3S$ and $LaH_{10}$.*

**Keywords:** *Generalized BCS equations for a superconductor subject to an applied field, Chemical potential, Characteristic parameters of a superconductor and their dependence on chemical potential*

## 1. Introduction

In a recent paper [1], hereafter referred to as I, it was shown that the chemical potential ($\mu$)-incorporated generalized BCS equations (GBCSEs) obtained from a Bethe-Salpeter equation (BSE) provide a unified framework for dealing with the $T_c$s, gaps and the coherence lengths of both elemental and high-$T_c$ superconductors (SCs). In the present paper we show that a similar approach is also applicable when one deals with the properties of such SCs in the presence of a magnetic field and when an electric current flows through them. The SCs dealt with here are Cd, Zn, Al, In, Hg and the high-$T_c$ SCs $MgB_2$, YBCO, Bi-2212, Bi-2223, Tl-2212, Tl-2223, and compressed $H_3S$ and $LaH_{10}$. While 10 of these SCs are the same as were studied in I, the remaining three (Zn, In and Hg) have been chosen in lieu of (Sn, Pb and Nb). This is so because the results of the study being carried out here for the latter SCs are already available in [2].

---

[#]Present address: Gurugram

*Corresponding author

For the convenience of introducing our notations, given below is a list of parameters that characterize an SC.

$T_c$: Critical temperature
*CPs*: Cooper pairs
$t = T/T_c$: Reduced temperature
$\Delta$: Gap of an elemental SC
$\Delta_1 < \Delta_2 < \Delta_3$: Gaps of a composite SC with three gaps
$\theta$: Debye temperature
$W$: The binding energy of a CP
$E_F$, $v_F$: Fermi energy, Fermi velocity
$\lambda$: Interaction parameter in the pairing equation due to the Coulomb repulsion between electrons and the attraction due to the ion-lattice
$m^* = \eta\, m_e$: Effective mass of an electron, $m_e$ being the free electron mass
$\xi$: coherence length at $T = 0$
*sf:* Self-field, the field that exists in the absence of any applied field
$H_c$: Critical field of an elemental SC
$H_{c1}$, $H_{c2}$: Lower and upper critical fields of a type II SC
$\lambda_m$: Magnetic interaction parameter in the pairing equation for an SC in an applied field
$N_L$: Landau index, i.e., the number of occupied levels when the (a, b) components of momentum are quantized when the SC is subject to an applied field in the c-direction
$\lambda_L$: London penetration depth at $T = 0$
$\kappa = \lambda_L/\xi$: Ginzburg-Landau parameter
$\mu$: Chemical potential
$N$: Demagnetization factor
$n_s$: Number density of charge-carriers
$v_c$: Critical velocity of Cooper pairs
$j_c$: Critical current density

Appealing to the empirical values of $T_c$ and $H_c/H_{c2}$, $\lambda_L$ and $j_c$ of any SC at $t = 0$, the objective of the present paper is to calculate the values of the following parameters related to them: $n_s$ and $v_c$ at $t \approx 0$; and $\lambda_m$, $N_L$ and, $\mu$ at both at $t \approx 1$ and 0.

The paper is organized as follows. In Section II, the $\mu$-incorporated GBCSEs recalled from earlier papers are recast in the form employed here. The applications of these equations are taken up in Section III. Unless stated otherwise, the units employed are Gaussian. The final sections sum up this study.

## 2. The $\mu$-Incorporated GBCSE

Since this paper is based on equations that have been derived in [2] and [3], the following succinct account of them is included for the sake of completeness. Our starting point is the parent BSE [4]

$$1 = \frac{V}{(2\pi\hbar c)^3}\frac{1}{2}\int_{\mu-k\theta}^{\mu+k\theta} d^3\boldsymbol{p}\,\frac{\tanh\left[\frac{1}{2k\,T}\left(p^2/2m^*-\mu-W/2\right)\right]}{\left(p^2/2m^*-\mu-W/2\right)}, \qquad (1)$$

$V/(2\pi\hbar c)^3$ plays the role of a propagator and $V$ - which is non-zero only in the range of integration - is the same parameter as occurs in *[N(0)V]* in the BCS theory, and $W$ is to be identified with $\Delta$ [5].

### 2.1 The GBCSEs for $H_c(t)$ or $H_{c2}(t)$ and the Number Density of Charge Carriers

Since subjecting an SC to an applied field considerably lowers the value of the magnetic interaction parameter $\lambda_m$ in the pairing equation, it turns out that one can employ the 1-phonon exchange mechanism (1PEM) for both the elemental and composite SCs without violating the Bogoliubov constraint that $\lambda_m$ must be positive and less than 0.5. However, since a high-$T_c$ SC has more than one ion-species to which the 1PEM may be attributed, we need to have a criterion for choosing one over the other. Although it is true that one can obtain the same values of $n_s$, $v_F$ and $\lambda_{L0}$ by choosing any of the constituent ion-species of such SCs, the values of $\mu_1$ and $N_{L1}$ corresponding to them are invariably different. We then need to invoke the requirement that $\lambda_{L0}$ must be greater than $\lambda_{L1}$; if this condition is satisfied by more than one species then, in principle, both of them are admissible candidates for the 1PEM and we need to appeal to experiment to find the values of $\mu_1$ to settle the issue.

The GBCSE of an SC in the presence of a magnetic field can be obtained from (1) by subjecting it to the Landau quantization scheme which replaces the total energy of the electron by

$$p_z^2/2m^* + \hbar\Omega_1(h)(n+1/2),$$

where $h \equiv H/H_c$ is the reduced applied field. Thus

$$\int d^3p \rightarrow \int dp_z \sum_n$$

and the $(p_x, p_y)$ degrees of freedom are quantized when the applied field is in the $z$-direction. The distribution of the total energy between the $(p_x, p_y)$ and the $p_z$ degrees of freedom is governed by the law of the equipartition of energy. Making these replacements in (1), employing $\xi = p_z^2/2m^* - \mu$, followed by $x = \xi/2ktT_c$, putting $\mu = q\rho k\theta$ with $\rho$ as a free parameter, and putting $W = 0$, we obtain the equation for $h_c$ or $h_{c2}$ at any $t = T/T_c$ between 0 and 1 as [3]

$$\text{E1}(t, h_c, \rho, q, \lambda_m) \equiv 1 - \lambda_m F(t, h_c, \rho, q) = 0, \qquad (2)$$

where

$$\lambda_m = \frac{ehH_cV}{16\pi^2}\sqrt{\frac{2\eta m_e}{q\rho k\theta}}$$

$$F(t, h_c, \rho, q) = \int_{L_-}^{L_+} \frac{dx}{\sqrt{1+\dfrac{2tT_cx}{\rho q\theta}}} \sum_{n=0}^{N_{L1}(..)} \frac{\tanh\left[x+(n+1/2)\dfrac{\hbar\Omega_1(h_c)}{2ktT_c}\right]}{\left[x+(n+1/2)\dfrac{\hbar\Omega_1(h_c)}{2ktT_c}\right]}$$

$$L_\pm = \frac{k\theta}{3}(-q\rho \pm 1),\ N_{L1}(..) = floor\left[\frac{2}{3}\frac{(\rho q+1)k\theta}{\hbar\Omega_1(h)} - \frac{1}{2}\right]$$

$$\Omega_1(h_c) = \frac{\Omega_0 h_c H_c}{\eta},\ \Omega_0 = 1.7588\ \text{s}^{-1}G^{-1}$$

Remarks:

a) For the employment of (2), the reduced field $h$ must be $\neq 0$, which of course is always so even when the applied field is 0 because then the self-field due to $j_c$ comes into play.

b) For the elemental SCs, we employ (2) at two values of $t$, viz 0.95 and 0.1. The former in lieu of 1 because at $t = 1$, $h = 0$, and the latter in lieu of 0 because $h_c$ can never be determined empirically at exactly $t = 0$ because the listed value of each of these at $t = 0$ is in fact a value extrapolated from a value of $t$ close to 0 for which the choice 0.1 seems to be reasonable. For the sake of convenience, for the elemental SCs, we label $t = 0.95$ as $t_1$ and $t = 0.1$ as $t_0$ because the former is close to 1 and the latter to 0; however, for the high-$T_c$ SCs, while employing for $t_1$ the same value as for the elemental SCs, we employ $t_0 = 0.05$ which is more realistic.

c) The substitution $\mu = q\rho k\theta$ in obtaining (2) enables us to employ this equation both at $t = t_1$ and $t_0$ because $\mu(t_1) \equiv \mu_1$ is parametrized as $\rho\, k\, \theta$ and $\mu(t_0) \equiv \mu_0$ as $q\, \mu_1 = q\, \rho\, k\, \theta$.

Labelling the values of $\lambda_m$ at $t = t_1$ and $t_0$ as $\lambda_{m1}$ and $\lambda_{m0}$, respectively, the definition of $\lambda_m$ above leads to

$$\lambda_{m0} = \lambda_{m1} \frac{h_{c0}}{h_{c1}} \frac{1}{\sqrt{q}}. \tag{3}$$

It follows from the above that $\lambda_{m1}$ can be obtained by solving (2) at $t = t_1$ with an assumed value of $\rho$ and $q = 1$. For the value of $h_c$ or $h_{c2}$ needed for this purpose, we employ the frequently used phenomenological Gorter-Casimir relation obtained via the 2-fluid model as

$$H_c(t), H_{c2}(t) = H_c(0), H_{c2}(0)[1-t^2], \tag{4}$$

where $H_c(0)$, $H_{c2}(0)$ denote the listed empirical values of these parameters. With the value of $\lambda_{m1}$ thus determined, we can solve (2) at $t = t_0$ by employing (3) for $\lambda_{m0}$ and the listed value of $H_c(0)$ or $H_{c2}(0)$ to obtain the value of $q$.

In order to fix $\rho$ - which is as yet a free parameter –we first employ the value assumed for it above together with the value of $q$ that it led to and calculate $n_s$ via the following number equation [6]

$$n_s(..) = C_1 (hH_c)^{3/2} \int_0^{\sqrt{L(..)}} \sum_{n=0}^{N_{L2}} F_n(..)dz, \; (h \neq 0) \tag{5}$$

where

$$C_1 = 2.1213\text{x}10^9, \;\; L(..) = \frac{1}{3}\frac{(\rho q+1)k\theta}{\hbar\Omega_1(h_c)} \;\; , \; N_{L2} = floor\left[\frac{2}{3}\frac{(\rho q+1)k\theta}{\hbar\Omega_1(h_c)} - \frac{1}{2}\right]$$

$$F_n(..) = \left[1 - \tanh\left\{\frac{\hbar\Omega_1(h)}{2ktT_c}\left(n + \frac{1}{2} + z^2 - \frac{q\rho k\theta}{\hbar\Omega_1(h)}\right)\right\}\right]$$

We are finally enabled to calculate λL via

$$\lambda_L = \sqrt{\frac{\varepsilon_0 m^* c^2}{n_s e^2}}, \text{ (S.I. units)} \tag{6}$$

where the permittivity of free space, $\varepsilon_0 = 8.85 \text{ x } 10^{-12}$ F/m and $e$ is the electronic charge. If the value of $\lambda_L$ thus obtained does not match its listed value, then we repeat the above procedure by varying $\rho$ till it does. Knowledge of $\rho$ and $q$ fixes $\mu_0$ – vide Remark c) after (2). We are hence enabled to calculate $v_F$ via

$$v_F = \sqrt{\frac{2\mu_0}{\eta m_e}}\, c \;(\mu_0, m_e \text{ in electron-Volt}) \tag{7}$$

**2.2 The GBCSE for $j_c(t, h)$**

This equation also follows from (1) and, for the reason noted for $H_c(t)$ or $H_{c2}(t)$, is required only in the 1PEM scenario. Essentially, the propagator $V/(2\pi\hbar c)^3$ is now non-zero only in the range

$$\mu_0 - k\theta \le \frac{(\boldsymbol{P}/2+\boldsymbol{p})^2}{2m^*},\ \frac{(\boldsymbol{P}/2-\boldsymbol{p})^2}{2m^*} \le \mu_0 + k\theta$$

where $\boldsymbol{P}$ is the momentum of a CP in the lab frame and $\pm\, \boldsymbol{p}$ the momentum of the constituents of the pair in the center-of-mass frame.

Employing the above propagator, two equations were derived in [3] – one for the situation where $\boldsymbol{P} = 0$ and hence $j_c(t, h) = 0$, and the other where these parameters are $\neq 0$. The solution of the former equation for any value of $\rho$ (which determines $\mu_1$) led to the value of $\lambda_{m1}$ which fixed $\lambda_{m0}$ via (3). The solution of the other equation then led to the value of $q$ which determines $\mu_0$. In this paper we have combined these equations into the following single equation

$$\mathrm{E2}(t,h,\rho,q,y) \equiv 1 - \frac{\lambda_m}{\sqrt{q}}\frac{h}{h_{c1}} J(t,h,\rho,q,y) = 0, \tag{8}$$

where

$$J(..) = \int_{z_-}^{z_+} dz \sum_{n=0}^{N_{L3}} \frac{\tanh(A_+) + \tanh(A_-)}{\left[z^2 - 1 + (n+\frac{1}{2})\frac{\hbar\Omega_1(h)}{q\rho k\theta}\right]},\ z_\pm = \sqrt{\frac{q\rho \pm 1 + 1/y}{3q\rho}}$$

$$N_{L3} = floor\left[\frac{2k\theta}{3\hbar\Omega_1(h)}\left(1 + q\rho - \frac{1}{y}\right) - \frac{1}{2}\right]$$

$$A_\pm = \frac{\rho k\theta}{2tT_c}\left(z^2 - 1 + (n+\frac{1}{2})\frac{\hbar\Omega_1(h)}{\rho q k\theta} \pm \frac{1}{\rho q y}\right)$$

$$y = k\theta/\alpha \;\; (\alpha = |\boldsymbol{P}||\boldsymbol{p}|\cos(\boldsymbol{P}.\boldsymbol{p})/2m^*).$$

The critical velocity $|\boldsymbol{v}_c|$ required to calculate $j_c = e\ n_s\ v_c$ is obtained via the dimensionless construct $y$ by the following equation

$$|\boldsymbol{v}_c| = \frac{c}{2y}\sqrt{\frac{6k\theta}{\rho q \eta m_e c^2}}. \tag{9}$$

Solving (2) by putting $t = t_m$, $h = h_m$, $q = 1$ and $1/y = 0$, where $t_m$ and $h_m$ are, respectively, the reduced temperature and reduced applied field at which $j_c$ is measured, we obtain the value of $\lambda_{m1}$, i.e., the magnetic interaction parameter corresponding to the situation when $\mathbf{v}_c = 0$ (and hence $j_c = 0$). Since $\lambda_{m0}$ is then fixed via (3), we can find the values of $q$ and $y$ corresponding to any value of $\rho$ by simultaneously solving the following equations

$$\mathrm{E2}(t_m,\ h_m,\ \rho,\ q,\ y)$$

$$\mathrm{E3}(t_m, h_m, \rho, q, y) \equiv 1 - \frac{e n_s(...) v_c(...)}{j_c(\exp)} = 0, \tag{10}$$

where $n_s$ and $v_c$ are given by (5) and (9), respectively. As will be seen below, the solutions of these equations lead to the values of several parameters in the list noted at the beginning of Section 1.

## 3. The Applications the $\mu$-Incorporated GBCSE

### 3.1 Calculation of Various Parameters Related to the Empirical Values of $\lambda_L$ of the Elemental SCs

We recall that while dealing with the $\xi$s of the elemental SCs in I, we had simultaneously solved two equations to obtain the values of $v_F$ and $q$ corresponding to an assumed values of $\rho$, which was varied till we were led to the listed values of the $\xi$s of these SCs. We found such a procedure to be impractical in dealing with the $\lambda_L$ of an elemental SC because it required an inordinately long time, due predominantly to the values of its Landau indices which are as large as of the order of $10^7$. Hence, the procedure followed by us for these SCs, as was outlined above, is:

Employing a guess value of $\rho$,

(i) Solve $\mathrm{E1}(t_1, h_{c1}, \rho, 1, \lambda_m)$ to determine $\lambda_{m1}$ at $t_1$ (= 0.95) and $h_{c1} = 0.0975$, where the latter value is obtained via (4).

(ii) With $t_0 = 0.1$ and $hc_0 = 0.99$ via (4), solve $\mathrm{E2}(t_0, h_{c0}, \rho, q, \lambda_{m1})$ for $q$. Hence $\mu_0 = \rho\, q\, k\, \theta$ and $\lambda_{m0}$ given by (3) are known. The solutions of the equations that determine $\lambda_{m1}$ and $q$ also yield the values of $N_{L1}$ and $N_{L0}$, which are the numbers of occupied Landau levels at $t = t_1$ and $t_0$, respectively.

(iii) Employ the above values of ($\rho$, $q$) to find $n_s$ at $t = t_0$ via (5).

(iv) Find $\lambda_L$ via (6).

Repeat the above sequence of steps by varying $\rho$ till the value of $\lambda_L$ it leads to matches the listed value of the latter. Although also laborious, compared with the procedure of solving two simultaneous equations, it takes much less time to obtain the final solution.

Remarks:

(i) We have assumed above that the value of $h_{c0}$ listed as pertaining to $t = 0$ is, in fact, the value that corresponds to $t_0$ (i.e., 0.1) because: (a) as was remarked earlier, it is obtained by extrapolation from its value at $t_0$ and (b) for values of $t < t_0$, the $h_c(t)$ vs. $t$ plot is nearly parallel to the $t$ axis. Statement (b) follows not only from (4), but also from several other such models as those of Baumgartner [7], Werthamer, Helfand and Hohenberg (WHH) [8], Jones, Hulm and Chandrasekhar [9] and Gor'kov [10]. A succinct account of these is given in a recent paper by Talantsev [11].

(ii) While the above remark justifies our identification of the *listed* values of $h_c$ at $t = 0$ with their values at $t_0$, it raises an important question posed below.

(iii) Question: What would be the value of $\lambda_L$ of an SC if we could empirically find $h_c(t)$ for $t \to 0$ by the method employed by Minkov et al., [12] for the hydrogen-rich high-$T_c$ SC? We draw attention to [13] for a discussion of this issue, following from which the answer to the question posed is: $\lambda_L$ may also exhibit a super-linear upswing as $t \to 0$, similar to the one known to be exhibited by $\xi$ as $t \to 1$.

For the calculation of $\lambda_L$ of any elemental SC, we adopt the same value of $\eta$ as was employed in I for the calculation of its $\xi$. The results of the above exercise for all the elements being dealt with are given in Table I which also includes the results for Sn, Pb and Nb that were reported in [2]. The values of $\theta$, $T_c$ and $H_c$ in Table 1 are taken from Poole [14].

**Table 1.** The calculated values of various parameters corresponding to the listed values of the $T_c$, $H_c$ and $\lambda_L$ of the elemental SCs studied in this paper. In column 3 are given the results obtained by solving Eq1(…) – vide (2) – with the inputs specified in column 2 together with $q$ =1, $t_1$ = 0.95, $h_{c1}$ = 0.0975; the value of $\rho$ in this column is one which leads, after trial and error, to a value close to $\lambda_L$(exp) as noted in column 5. The values of $q$, $\mu_0$ and $\lambda_{m0}$ in column 4 are also obtained by solving Eq1(…) with the input of $\lambda_{m1}$ together with $t_0$ = 0.1 and $h_{c0}$ = 0.99. The values of $n_s$, $v_F$ and $\lambda_L$(th) are obtained via (5), (7) and (6), respectively.

| SC | $\theta$ (K), $T_c$ (K)<br>$H_c$ (G), $\eta$<br>$\rho$, $\mu_1 = \rho k\theta$ (eV) | $\lambda_{m1}$<br>$N_{L1}$ | $q$, $\mu_0 = q\mu_1$ (eV)<br>$\lambda_{m0} = \frac{\lambda_{m1}}{\sqrt{q}} \frac{h_c(0.1)}{h_c(0.95)}$<br>$N_{L0}$ | $n_s$ ($10^{27}$ $m^{-3}$)<br>$v_F$ ($10^5$ m/s)<br>$\lambda_L$(th), $\lambda_L$(exp) (nm) |
|---|---|---|---|---|
| 1 | 2 | 3 | 4 | 5 |
| Cd | 210, 0.42,<br>30, 0.73<br>51.2, 0.9265 | 1.400.10$^{-7}$<br>1.358.10$^{7}$ | 1.7392, 1.611<br>1.077.10$^{-6}$<br>2.306.10$^{6}$ | 1.701<br>8.81<br>110.1, 110.0 |
| Zn | 309, 0.535<br>54, 0.85<br>210, 5.539 | 1.329.10$^{-7}$<br>5.174.10$^{7}$ | 1.6515, 9.147<br>1.050.10$^{-6}$<br>8.399.10$^{6}$ | 28.55<br>19.5<br>28.99, 29 |
| Al | 428, 1.16<br>105, 1.48<br>270, 9.958 | 1.112.10$^{-7}$<br>8.321.10$^{7}$ | 1.6819, 16.75<br>8.702.10$^{-7}$<br>1.376.10$^{7}$ | 162.3<br>20.0<br>16.04, 16.0 |
| In | 112, 3.41<br>225. 1.37<br>256, 2.471 | 1.383.10$^{-6}$<br>8.920.10$^{6}$ | 2.0575, 5.084<br>9.791.10$^{-6}$<br>1.804.10$^{6}$ | 24.16<br>11.4<br>40.0, 40.0 |
| Sn | 195, 3.72<br>305, 1.26<br>144, 2.420 | 1.138.10$^{-6}$<br>5.945.10$^{6}$ | 2.017, 4.881<br>8.128.10$^{-6}$<br>1.177.10$^{6}$ | 20.01<br>11.7<br>42.07, 42.0 |
| Hg | 72, 4.16<br>410, 1.88<br>319, 1.979 | 3.134.10$^{-6}$<br>5.377.10$^{6}$ | 2.198, 4.354<br>2.145.10$^{-5}$<br>1.163.10$^{6}$ | 30.7<br>9.02<br>41.56, 41.5 |
| Nb | 276, 9.25,<br>1580, 12<br>29.8, 0.709 | 5.589.10$^{-6}$<br>3.286.10$^{6}$ | 2.4227, 1.717<br>3.646.10$^{-6}$<br>7.691.10$^{5}$ | 125.2<br>2.24<br>52.02, 52.0 |

Table 1 (Continued)

| | | | | |
|---|---|---|---|---|
| Pb | 96, 7.2<br>800, 1.97<br>241.6, 1.999 | $4.720.10^{-6}$<br>$2.919.10^{6}$ | 2.329, 4.655<br>$3.141.10^{-5}$<br>$6.680.10^{5}$ | 36.5<br>9.12<br>39.03, 39.0 |

Since we could not find any data on the empirical values of $j_c$ of the elemental SCs, presumably because they are of no practical interest, we now move on to the high-$T_c$ SCs.

### 3.2 Calculation of Various Parameters Related to the Empirical Values of $\lambda_L$ and $j_c$ of the High-$T_c$ SCs

Insofar as $\lambda_L$ is concerned, the procedure followed for obtaining the values of various parameters related to it for any high-$T_c$ SC is similar to the one followed for the elemental SCs, except that (a) we let $t_0 = 0.05$ which is more in accord with the lowest temperature at which data are now becoming available rather than 0.1, as is exemplified by $H_3S$ ($T_c$ = 196) for which $\lambda_L$ has been reported in [12] at 10 K and for which therefore $t_0 = 0.05$, and (b) as elaborated below, we need to choose the ion-species in these composite SCs to which the empirical value of $\lambda_L$ may be attributed.

In $MgB_2$, CPs can result via the 2PEM where the electrons are bound via simultaneous phonon-exchanges due to the B and Mg ions, as well via the 1PEM where the electrons are bound due to the B and the Mg ions individually. Since we require the 1PEM for dealing with the $\lambda_L$ of the SC, we need to choose between the B and the Mg ions. In this case the natural choice is the B ions because the action of the Mg ions is due to the proximity effect. Similarly, in dealing with YBCO we need to choose between the Y and the Ba ions both of which can individually cause pairing. However, there is now no natural criterion for preferring one ion-species over the other because – by appropriately choosing $\rho$ - the desired value of $\lambda_L$ can be obtained by invoking either of these ion-species. While, unsurprisingly, the values of $N_{L0}$, $n_s$ and $v_F$ in both cases are found to have nearly the same values, the values of {$\mu_1$ (meV), $N_{L1}$, $N_{L0}$) are different. Since we need to have $N_{L0} < N_{L1}$, we must choose the ion-species which satisfies this condition. It is notable that this conclusion can be verified by monitoring $\mu_1$ via the Hall effect. While all the ion-species to which the 1PEM may be attributed were duly considered for each of the SCs in Table 2, the results given therein are only those that satisfy the $N_{L0} < N_{L1}$ criterion. We finally note that for the Bi-based SCs, there are three candidates for the 1PEM, viz., the ion-species of Bi, Ca and Sr and that for the super-hydrides, akin to $MgB_2$, the natural choice is that of the H ions. In Table 2, the values of the empirical parameters of $MgB_2$ have been taken from [15] and those of the cuprates and the hydrides from [14] and [12], respectively.

**TABLE 2**. The calculated values of various parameters corresponding to the listed values of $\lambda_L$ of the high-$T_c$ SCs studied in this paper. For the values of the $\theta$s and $\eta$ (excepting $LaH_{10}$) in column 1, see I; for LaH10, see Remark (i) below the Table. In column 3 are given the results obtained by solving Eq1(…) – vide (2) – with the inputs specified in column 2 together with $q$ =1, $t_1 = 0.95$, $h_{c1} = 0.0975$, $\theta_{1P}$ being the value employed in the scenario of the 1PEM and chosen from among the $\theta$-values in column 1; the value of $\rho$ in this column is one which leads, after trial and error, to the value of $\lambda_L$(exp) as noted in column 5. The values of $q$ and $\mu_0$ in column 4 are also obtained by solving Eq1(…) with the input of $\lambda_{m0}$ together with $t_0 = 0.05$ and $h_{c0} = 0.9975$. The values of $n_s$, $v_F$ and $\lambda_L$(th) are obtained via (5), (7) and (6), respectively. $MgB_2$/B means that the 1PEM in this SC is being attributed to the B ions, and likewise for all the other SCs.

| SC<br>$\theta_{SC}$ (K)<br>$\theta_1$ (K)<br>$\theta_2$ (K)<br>$\eta$ | $\theta_{1P}$ (K), $T_c$ (K), $H_{c2}$ (T)<br>$\rho,\ \mu_1 = \rho k\theta$ (meV) | $\lambda_{m1}$<br>$N_{L1}$ | $q, \mu_0 = q\mu_1$ (meV)<br>$\lambda_{m0} = \frac{\lambda_{m1}}{\sqrt{q}} \frac{h_c(0.05)}{h_c(0.95)}$<br>$N_{L0}$ | $n_s$ ($10^{27}$ $m^{-3}$)<br>$v_F$ ($10^5$ m/s)<br>$\lambda_L$ (th) (nm)<br>$\lambda_L$ (exp) (nm) |
|---|---|---|---|---|
| 1 | 2 | 3 | 4 | 5 |
| $MgB_2$/B<br>θ=815<br>$\theta_B$=1062<br>θMg=322<br>0.44 | 1062, 40, 2.5<br>3.695, 338.2 | 8.190.10<br>-5<br>4465 | 3.886, 1316<br>$4.250.10^{-4}$<br>1427 | 0.634<br>10.3<br>139.9<br>140 |
| YBCO/Y<br>θ=410<br>$\theta_Y$=410<br>θBa=117<br>3.0 | 410, 90, 98<br>1.80, 63.6 | 2.768.10<br>-3<br>178 | 12.276, 780.7<br>$8.081.10^{-3}$<br>143 | 5.00<br>3.03<br>130.1<br>130 |
| Bi-2212/Sr<br>θ=237<br>$\theta_{Ca}$=237<br>$\theta_{Bi}$=269<br>$\theta_{Sr}$=286<br>2.75 | 286, 86, 36<br>1.00, 24.65 | 1.827.10<br>-3<br>221 | 13.28, 327.3<br>$5.189.10^{-3}$<br>154 | 1.24<br>2.05<br>250.2<br>250 |
| Bi-2223/Sr<br>θ=275<br>$\theta_{Ca}$=275<br>$\theta_{Bi}$=312<br>$\theta_{Sr}$=331<br>3.0 | 331, 110, 58<br>1.008, 28.75 | 2.660.10<br>-3<br>174 | 17.84, 512.8<br>$6.443.10^{-3}$<br>161 | 2.71<br>2.45<br>176.8<br>177 |
| Tl-2212/Ba<br>θ=254<br>$\theta_{Ca}$=254<br>$\theta_{Tl}$=289<br>$\theta_{Ba}$=296 | 296, 100, 100<br>1.02, 26.02 | 5.345.10<br>-3<br>90 | 19.05, 495.6<br>$1.253.10^{-2}$<br>89 | 2.55<br>2.41<br>182.3<br>182 |
| Tl-2223/Ba<br>θ=290<br>$\theta_{Ca}$=290<br>$\theta_{Tl}$=330<br>$\theta_{Ba}$=338 | 338, 121.5, 115<br>1.005, 29.27 | 5.447.10<br>-3<br>89 | 19.64, 575.0<br>$1.257.10^{-2}$<br>90 | 3.19<br>2.60<br>162.9<br>163 |
| $H_3S$/H<br>θ=1531<br>$\theta_H$=1983.2<br>$\theta_S$=174.5<br>2.76 | 1983.2, 196, 97<br>11.4, 1948.3 | 3.065.10<br>-4<br>3560 | 4.5417, 8848<br>$1.471.10^{-3}$<br>1481 | 163<br>10.6<br>21.9<br>22.0 |
| $LaH_{10}$/H<br>θ=1156<br>$\theta_H$=1248.6<br>$\theta_{La}$=1146.7<br>3.0 | 1248.6, 231, 143.5<br>8.1, 871.5 | 8.716.10<br>-4<br>1208 | 6.4691, 5638<br>$3.506.10^{-3}$<br>692 | 93.7<br>8.13<br>30.1<br>30 |

Remarks:

(i) The value of $\eta$ for $LaH_{10}$ in I was assumed to be the same as for $H_3S$, i.e., 2.76. Following Balbaa and Manchester [16], it has been revised to 3.0 in this paper.

(ii) In the above Table, for $MgB_2$ we have $v_F$ =10.3 x $10^5$ m/s corresponding to $\eta$ = 0.44 and $\lambda_L$ = 140 nm, whereas for the same value of $\eta$ and $\xi$ = 8.1 nm in I it was 2.7 x $10^5$ m/s. It is notable here that besides 0.44, a multitude of other values of $\eta$ for this SC have also been reported in the literature, which – see I for references - are: 0.44 - 0.66, 1.25, 1.1, 0.47, 0.50 and 1.08 - 1.20. If, in the light of these, the value of $\eta$ is changed from 0.44 to 1.1 (because it has been reported in more than one paper) then, following the same procedure as for $\eta$ = 0.44, we end up with, for $\rho$ = 2.5, $\lambda_L$ = 167.7 nm and $v_F$ = 4.8 x $10^5$ m/s. Interestingly, the value of $\lambda_L$ still falls in the range (85 – 180) nm given in (15), and that of $v_F$ is now in accord with its value given by Leggett [17] as ≈ 5 x $10^5$ m/s.

(iii) The values of $v_F$ of $H_3S$ and $LaH_{10}$ in the above Table are 10.6 x $10^5$ and 8.13 x $10^5$ m/s, respectively, whereas the value employed for both of them in I was 2.5 x $10^5$ m/s. The basis for the last of these values was the range of $v_{Funiversal}$ for the cuprates given by Talantsev [18] as ≈ (2.5 – 3.8) x $10^5$ m/s, which follows from the equation

$$v_{F,univ} \approx (2\Delta / 1.3kT_c)\text{x}10^5 \text{ m/s}$$

derived by him. We note that in order to obtain a *universal* upper bound on the value of $v_F$ we must use the lowest value of $T_c$ and the highest value of Δ, which in the present instance are $T_c$ = 196 K for $H_3S$ and - as derived by Kruglov et al., [19] - Δ = 62 meV for $LaH_{10}$. These values raise Talantsev's upper bound on $v_{F,\,univ}$ to 5.6 x $10^5$ m/s with which the aforesaid values of $H_3S$ and $LaH_{10}$ are in reasonable agreement considering the number of parameters and the diversity of relations among them that have been employed.

(iv) It seems remarkable that the values of both $\eta$ and $v_F$ – pertaining to the averaged values $\xi$ in I and those of $\lambda_L$ of all the high-$T_c$ SCs barring the three discussed above - were found to be the nearly the same.

(v) The 1PEM in YBCO has been attributed to the Y ions because while the choice of the Ba ions also led to almost the same values of $n_s$, $v_F$ and $\lambda_L$ (via $\rho$ = 2.8 and $q$ = 28.59) as the Y ions, the values of $N_{L1}$ = 68 and $N_{L0}$ = 143 that it led to did not satisfy the criterion that $N_{L0}$ must be *less* than $N_{L1}$.

(vi) Following Ramshaw et al., [20] where η ≈ 3 has been reported as the representative of the cuprate family of SCs, we had adopted the value η = 3 in I for all such SCs. In this paper too we have followed this practice for all the cuprates excepting Bi-2212, for which we found that in the 1PEM scenario, none of the ion-species led to the listed value of $\lambda_L$ if $\eta$ = 3 exactly. Upon employing $\eta$ = 2.75 we found that only the Sr ions led to the value of $\lambda_L$ being sought. As for the Bi and the Ca ions, we found that even with $\eta$ = 2.75, for $\lambda_m$ to remain real, the maximum values of $\lambda_L$ that they led to were 243.4 and 235.4 (nm), respectively, each of which falls short of $\lambda_L$ = 250 nm being sought.

(vii) It is notable that the value of $q$ – let's call it $q_c$ - for any SC reported in column 4 of the above Table is not a unique solution of E1(….) - vide (2). In fact, a plot of this equation for values of $q$ slightly less than and slightly more than $q_c$ comprises a band of many zig-zag lines that extends diagonally from below the $q$ = 0 axis to above it. Among these are several lines that cross the $q$ = 0 axis before the band moves away from it. Each such crossing marks a root of the equation. In other words, the equation we are solving has multiple roots. However, in order to specify $\lambda_L$ and the other related parameters to the warranted accuracy of the problem we are addressing, it turns out that the choice of $q_c$ from among such values is generally immaterial because they constitute a closely packed bunch. For a graphical representation of the foregoing account, we draw attention to Figure 1 in [21].

(viii) We finally note that the solutions of Eq1(…) yielding multiple, oscillatory values of $q$ are reminiscent of the oscillations in the values of the magnetization or the magnetic susceptibility of a material when plotted against the applied field $H_m$ and are well known as the de Haas-van Alphen effect. That the $T_c(H)$ plot for an SC should also exhibit similar behavior was pointed out, perhaps for first time, by Gunther and Gruenberg [22]. Returning to the present study and considering the totality of the solutions of Eq1(…) rather than only $q_c$, the oscillatory behavior of $q$ being reported here signifies that as the chemical potential varies due to temperature and/or the applied field, each one of the following parameters must also exhibit the de Haas-van Alphen effect: $n_s$, $v_F$, , $N_L$ and $\lambda_L$.

We now move on to deal with the listed values of $j_c$ of the high-$T_c$ SCs, noting at the outset that the framework for their calculation in this paper is essentially the same as was employed in [3] for Bi-2212. To put in perspective the difference between the present and our earlier treatment of this and similar SCs, we note that, typically, the $j_c$ of an SC is listed as: $j_c$, the applied field ($H_m$), and the temperature ($T_m$) at which $j_c$ is measured, as in [14].

Missing alongside the three entries for $j_c$ is another parameter, viz., $\eta$, which is needed in our framework because its basic equation is $j_c = e\, n_s\, v_c$, and $v_c$ depends on $\eta$. Briefly, the procedure for calculating $j_c$ that we had followed in [3] comprised the following steps after assuming a value of $\eta = \eta_1$ (say): (a) assume a value of $\rho$ (which fixes $\mu_1$) to find $\lambda_{m1}$ via Eq2(…) with $q = 1$, $1/y = 0$ and the other requisite inputs, (b) employ these values of $\rho$ and $\lambda_{m1}$ to find $q$ (which fixes $\mu_0$) and y by simultaneously solving Eq2(…) and Eq3(…) with $\eta = \eta_2, j_c = j_c$(exp) and the other requisite inputs. The values of $q$ and $y$ then enabled us to calculate each member of the set $S = \{\mu_1, \mu_0, \lambda_{m0}, v_c, n_s, v_F, N_{L1}$ and $N_{L0}\}$ corresponding to $j_c$(exp). Note that we had allowed $\eta$ to have different values when $q = 1$ (which corresponds to $j_c = 0$) and when $q \neq 1$ (which corresponds to $j_c \neq 0$). With this assumption, even after restricting $\eta_1$ and $\eta_2$ to within reasonable limits, we ended up with the result that innumerable values of the triplet $T_r = \{\rho, \eta_1, \eta_2\}$ and therefore of the set $S$ (…) can account for any *single* value of $j_c$(exp), whereas reported herein corresponding to it are *unique* values of the triplet $T_r$ (…) and the set $S$ (…).This refinement has come about because of the reasons discussed below.

Extensive survey of the relevant literature suggested to us that: (i) adopting different values of $\eta$ corresponding to $j_c = 0$ and $\neq 0$ is unwarranted; (ii) as discussed in I, it is reasonable to adopt a single value of $\eta$ as $\approx 3$, 0.56, 2.76 and 3.0 for the cuprates, $MgB_2$, $H_3S$ and $LaH_{10}$, respectively, and (iii) relying on the widely accepted bounds on the values of $v_F$ of each SC, such as $(2.7 \pm 0.5) \times 10^5$ m/s for the cuprate family of SCs, we can find a unique value of $\rho$ – and therefore of each member of the set $S$ (…) – which leads to the value of $v_F$ in conformity with these bounds.

There is another, important, feature that comes into play when the field $H_m$ at which $j_c$ is measured is zero, as is the case for the Tl-based SCs and the hydrides in Table 3. In this case, in lieu of $H_m (T_m)$ we need to employ $H_p (T_m)$ which marks the highest value of the self-field beyond which $j_c (T_m)$ is zero. The calculation of $H_p (T_m)$ involves the following steps:

(i) Calculate $H_{c1}(0)$ via the well-known result

$$H_{c1}(0) = H_{c2}(0)\frac{\ln(\kappa)}{2\kappa^2}.$$

(ii) Calculate the demagnetization factor $N$, which for a rectangular cuboid with dimensions of width $w$ and thickness $th$, is given by [23]

$$N \approx 1 - \frac{2th}{\pi w}\ln\left(\frac{2w}{th}\right).$$

(iii) Calculate $H_p (0)$ via

$$H_p(0) = (1 - N)H_{c1}(0).$$

(iv) Calculate $Hp\ (T_M)$ via

$$H_p(T_m) = H_p(0)\left[1-(T_m/T_c)^2\right].$$

The value of $H_p\ (T_m)$ for Tl-2212 (Tl-2223) is obtained by adopting the following values for the requisite parameters: $T_c$ = 100 K (121.5 K), $H_{c2}(0)$ = 100 T (115 T) –which are in accord with their values in Table 2; $\kappa$ = 93.8 (161.4) due to its value of $\lambda_L$ given in Table 2 and that of its $\xi$ in I; $H_{c1}$ = 258.0 G (112.2 G), $N$ = 0.93 (0.81), $Hp(0)$ = 17.68 G (21.33 G), $H_p(T_m)$ = 7.20 G (12.76 G) - all of these on the basis of the above equations and the following specifications [24] of the sample the $j_c$ values of which is being addressed: $j_c$ = 1 x $10^6$ A/cm$^2$ (7 x $10^5$ A/cm$^2$), $w$ = 18 μm (18 μm), $t$ = 440 nm (1790 nm), N = 0.93 (0.81), ($T_m$ = 77 K (77 K). The values of the required parameters for the other cuprates in Table 3 are taken from [14] and those of $MgB_2$ and the super-hydrides from [15] and [25], respectively. The results pertaining to the $j_c s$ of all the SCs being dealt with given in Table 3 correspond to values of $\rho$ which – as elaborated in I - lead to $v_F$ = 2.7 x $10^7$ cm/s for both $MgB_2$ and the cuprates, and to 3.1 x $10^7$ cm/s for the super-hydrides.

**Table 3.** The calculated values of various parameters corresponding to the listed values of $j_c$ of the high-$T_c$ SCs studied in this paper. For the values of the $\theta$s and $\eta$ in column 1, see I. In column 3 are given the results obtained by solving Eq2(…) – vide (2) – with the value of $\rho$ specified in column 2 together with $q$ =1, $1/y$ = 0 (signifying that $j_c$ = 0, which is the reason $\lambda_m$ has now been labelled not as $\lambda_{m1}$, but as $\lambda_m(j_c{=}0)$), $t_m = T_m/T_c$,, $h_m = H_m/H_{c2/p}(0)$ $\theta_{1P}$ being the value employed in the scenario of the 1PEM and chosen from among the $\theta$-values in column 1; the value of $\rho$ is one which leads, after trial and error, to a value of $v_F$ close to $v_F$ (exp) as noted in column 5. The values of $q$ and $y$ in column 4 are obtained by simultaneously solving Eq2(…) and Eq3(…) with the input of $\rho$, $\lambda_m(j_c{=}0)$, $t_m$, $h_m$ and $j_c$(exp). The values of $n_s$, $v_c$, and $v_F$ in column 5 are obtained via (5), (9) and (7), respectively. $MgB_2$/B denotes that the 1PEM for this SC is being invoked via the B ions, and likewise for the other SCs.

| SC, $\theta_{SC}$(K)<br>$\theta_1, \theta_2$(K), $\eta$<br>$T_m$(K), $H_m$ or $H_p$<br>$j_c$(exp) (A/cm$^2$) | $\theta_{1P}$(K), $T_c$(K)<br>$H_{c2\ \mathrm{or}\ p}(0)$, $\rho$<br>$\mu(j_c=0)$<br>$=\rho k\theta_{1P}$ (meV) | $\lambda_m(j_c=0)$<br>$N_L(j_c=0)$ | $q$, $y$, $\mu(j_c)=$<br>$q\mu(j_c=0)$ (meV)<br>$\lambda_m(j_c)=$<br>$\frac{\lambda_m(j_c=0)}{\sqrt{q}}$<br>$N_L(j_c)$ | $n_s$(cm$^{-3}$)<br>$v_c$(cm/s)<br>$v_F$ (th) ($10^7$cm/s)<br>$v_F$ (exp) ($10^7$cm/s)<br>$j_c=en_sv_c$ (A/cm$^2$) |
|---|---|---|---|---|
| 1 | 2 | 3 | 4 | 5 |
| $MgB_2$/B,<br>$\theta$=815<br>$\theta_B$=1062,<br>$\theta_{Mg}$=322, 0.56<br>4.2, $H_m$=4 T<br>$1.10^6$ | 1062, 40<br>$H_{c2}(0)$ =14 T<br>1.7, 155.6 | $6.799.10^{-4}$<br>198 | 0.7387, 102.2<br>114.9<br>$7.190.10^{-4}$<br>165 | $3.44.10^{19}$<br>$1.81{\cdot}10^5$<br>2.69<br>2.70<br>$1.10^6$ |
| YBCO/Y,<br>θ=410<br>$\theta_Y$=410,<br>$\theta_{Ba}$=117<br>3.0<br>4.2, $H_m$=0.3 T<br>$1.2.10^6$ | 410, 90,<br>$H_{c2}(0)$ = 98 T<br>17.3, 611.2 | $2.487.10^{-5}$<br>$3.72.10^4$ | 0.9936, 627.8<br>607.3<br>$2.495.10^{-5}$<br>$3.70.10^4$ | $3.50.10^{21}$,<br>$2.14.10^3$<br>2.67<br>2.70<br>$1.2.10^6$ |
| Bi-2212/Sr<br>θ=237<br>$\theta_{Ca}$=237,<br>$\theta_{Bi}$=269<br>$\theta_{Sr}$=286<br>3.0<br>4.2, $H_m$=12 T<br>$1.10^6$ | 286, 89<br>$H_{c2}(0)$ = 36T<br>25.0, 616.1 | $1.470.10^{-3}$<br>922 | 0.9904, 515.6<br>610.2<br>$1.477.10^{-3}$<br>913 | $3.44.10^{21}$<br>$1.81.10^3$<br>2.67<br>2.70<br>$1.10^6$ |
| Bi-2223/Sr,<br>θ=275<br>$\theta_{Ca}$=275,<br>$\theta_{Bi}$=312<br>$\theta_{Sr}$=331<br>3.0<br>4.2, $H_m$=13 T<br>$1.10^5$ | 331, 110<br>$H_{c2}(0)$ = 58 T<br>22.0, 627.5 | $1.357.10^{-3}$<br>871 | 0.9959, 6153<br>624.9<br>$1.360.10^{-3}$<br>867 | $3.59.10^{21}$<br>$1.74.10^2$<br>2.71<br>2.70<br>$1.10^5$ |
| Tl-2212/Ba,<br>θ=254<br>$\theta_{Ca}$=254,<br>$\theta_{Tl}$=289<br>$\theta_{Ba}$=296<br>3.0<br>77, $H_m$=0<br>$H_p(T_m)$=7.20 G<br>$1.10^6$ | 296, 100<br>$H_p(0)$ = 1.77 mT<br>24.5, 624.9 | $1.762.10^{-7}$<br>$1.56.10^7$ | 0.9932, 543.1<br>620.7<br>$1.768.10^{-7}$<br>$1.55.10^7$ | $3.53.10^{21}$<br>$1.77.10^3$<br>2.70<br>2.70<br>$1.10^6$ |
| Tl-2212/Ba,<br>θ=254<br>$\theta_{Ca}$=254,<br>$\theta_{Tl}$=289<br>$\theta_{Ba}$=296<br>3.0<br>77, $H_m$=0<br>$H_p(T_m)$=7.20 G<br>$1.10^6$ | 296, 100<br>$H_p(0)$ = 1.77 mT<br>24.5, 624.9 | $1.762.10^{-7}$<br>$1.56.10^7$ | 0.9932, 543.1<br>620.7<br>$1.768.10^{-7}$<br>$1.55.10^7$ | $3.53.10^{21}$<br>$1.77.10^3$<br>2.70<br>2.70<br>$1.10^6$ |
| Tl-2212/Ba,<br>θ=254<br>$\theta_{Ca}$=254,<br>$\theta_{Tl}$=289<br>$\theta_{Ba}$=296<br>3.0<br>77, $H_m$=0<br>$H_p(T_m)$=7.20 G<br>$1.10^6$ | 296, 100<br>$H_p(0)$ = 1.77 mT<br>24.5, 624.9 | $1.762.10^{-7}$<br>$1.56.10^7$ | 0.9932, 543.1<br>620.7<br>$1.768.10^{-7}$<br>$1.55.10^7$ | $3.53.10^{21}$<br>$1.77.10^3$<br>2.70<br>2.70<br>$1.10^6$ |

Table 3 (Continued)

| | | | | |
|---|---|---|---|---|
| $H_3S$/H<br>θ=1531,<br>$\theta_H$=1983.2<br>$\theta_S$=174.5<br>2.76<br>100, $H_m$=0<br>$H_p\ (T_m)$=71 mT<br>7.0.10$^6$ | 1983.2, 196,<br>$H_p(0)$ = 96mT<br>3.13, 534.9 | 2.296.10$^{-6}$<br>1.58.10$^7$ | 0.9939, 540.4<br>531.7<br>2.302.10$^{-6}$<br>1.57.10$^5$ | 3.26.10$^{21}$<br>1.34.10$^4$<br>2.60<br>2.60<br>7.10$^6$ |

Remarks

i) Denoting the final solutions for any SC given above by the set Σ(SC) = {$\mu_0$, $N_{L0}$, $n_s$, $v_c$, $v_F$, $j_c$}, it is seen that Σ(YBCO) has been obtained by employing Y ions in the scenario of the 1PEM. Since the 1PEM for this SC can also be due to the Ba ions, we note that – with the same units as in the Table - Σ(YBCO/Ba) = {613.5, 35910, 3.35.10$^{21}$, 2.33.10$^3$, 2.68.10$^7$, 1.2.10$^6$} ≅ Σ(YBCO/Y); the major difference between the two cases is in the values of $\rho$, which are: 611.2 and 625.1 meV for the Y and Ba ions, respectively. Thus, the value of $j_c$ for YBCO can be attributed to either of these two ion-species.

ii) The above remark also applies to the Bi- and Tl-based SCs.

## 4. Discussion

In dealing with the $\xi$-values of the SCs in I, it was found that the values of $\mu_0$ for the elemental SCs were of the order of electronVolts and those of the high-$T_c$ SCs were of the order of milli electron Volts. It is notable that this feature persists even while dealing with the $\lambda_L$s and $j_c$s of the these SCs.

A striking feature of the Landau index $N_{L0}$ corresponding to the $\lambda_L$s of the elemental and high-$T_c$ SCs we have dealt with is: while it is of the order of $10^6$ or more for the former category, it is of the order of $10^3$ or less for the latter. The extent of the difference between the two categories of SCs on a scale extending from the classical to the extreme quantum behavior can be gauged by recalling Bohr's Correspondence Principle, which may be paraphrased as: for any phenomenon exhibited by any system, if the associated quantum number $N \rightarrow \infty$, then the behavior of the system will not be too different from that predicted by the classical laws, whereas if $N \rightarrow 1$, then it will be imperative to employ the quantum laws.

It was noted above that in the GBCSEs-based approach we need to invoke more than the 1PEM while dealing with the $T_c$ and the multiple gaps of a high-$T_c$ SC in order to comply with the Bogoliubov constraint. This is similar to invoking more than one band in the multi-band approach (MBA) for the same purpose. However, as was also noted above, while dealing with the properties of any such SC subjected to an applied field, we need to employ only the 1PEM. It is interesting to note that this feature too has a parallel in the MBA. This is evidenced by Audouard et al., [26] who have remarked that their study concerned with an iron-based SC "suggested that one band mainly controls the superconducting properties in magnetic fields despite the multiband nature of the Fermi surface." For a comparative study of the GBCSEs-based approach and the MBA, we draw attention to [27].

Notwithstanding the general perception that the effect of an external field is to adversely affect the superconducting properties of a system, it has been reported by Rasolt and Tešanović [28] that in systems with low carrier density for which $N_L$ is a small number, there comes about a strong enhancement in $T_c$ due to the enhancement of the effective electron-electron interaction with increasing magnetic field. We now draw attention to Table II, where for all the SCs are given the values of $\lambda_{m1}$ in column 3 and those of $\lambda_{m0}$ in column 4. For $MgB_2$, it is seen that in going from ($h_{c1}$= 0.0975, $N_{L1}$= 4470) to ($h_{c0}$= 0.9975, $N_{L1}$= 1429), $\lambda_{m1}$ = 8.2 x $10^{-5}$ goes over to $\lambda_{m0}$ = 4.2 x $10^{-4}$ – an increase by a factor of five. This is also a feature of the remaining SCs in Table II, though the values of $\lambda_{m0}/\lambda_{m1}$ for them are different - varying between 2.4 to 2.9 for the cuprates and 4.0 and 4.9 for the super-hydrides. It is hence seen that the GBCSEs-based approach not only corroborates the finding reported in [28] that the effective electron-electron interaction increases with increasing magnetic field, but also provides a quantitative estimate of it.

In dealing with $\lambda_L$ of the SCs, we needed to employ the GBCSEs in this paper at only two points, viz., one close to $t$ = 1 and the other close to $t$ = 0. It is notable therefore that at the expense of employing a model for the variation of $\mu$ for $0 \le t \le 1$, it has been shown [29] that they can also be employed to provide fits to the empirical values of $H_{c2}(t)$ which, up to the lowest value of $t$ for which such data are available for $H_3S$, are as good as those obtained in [11] by employing not only the WHH model, but also an ingenious mix of several others. From [29] and Table II in this paper, it is seen that the GBCSEs-based approach not only provides good fits to the empirical $H_{c2}$ data, but also leads to the values of several related parameters which shed light on such important features as have been discussed in the preceding two paragraphs.

## 5. Conclusion

We finally deal with the role that the GBCSEs-based approach followed here and in our earlier papers might play in the quest for clues to the fabrication of SCs with bespoke properties. It is notable in this connection that, typically – see [14], for example – the $j_c$-values of the high-$T_c$ SCs are reported under four heads: $j_c$ (A/cm$^2$), $B_{app}$ (T), $T_{meas}$ (K) and Comments (under which is given information of varied nature). For the study carried out in this paper, we additionally needed the values of $\theta$, $T_c$, $H_{c2}$, $v_F$, $\eta$ and the dimensions of the SC (which are sometimes given under the Comments column), several of which were taken from an assortment of sources. This is not an ideal situation. What we deem desirable is that (a) several samples of the parent SC the $j_c$ of which is of interest be fabricated differing from one another in size, shape, method of preparation and the nature of doping; (b) to the extent feasible, the above-named parameters of *each* of these samples be specified on an empirical basis; (c) the $j_c$s of all the samples be measured at the same values of $T_m$ and $H_m$ of practical interest and (d) subjected to a study similar to one the results of which have been given in Table III. Carrying out this exercise, we will end up with the values of $\mu$ at $t$ close to one and zero for each of these samples, besides those of various parameters dependent on them. It is plausible that our quest for optimizing $j_c$ culminates in the result that we need only to control this single parameter, i.e., $\mu$, which, interestingly, would suggest that, in a sense, it's role in SCs is similar to that of the codons in the strands of the DNA of living beings.

## References

[1] G. P. Malik and V. S. Varma, "On the Properties of Elemental and High-T$_c$ Superconductors in a Unified Framework I", World Journal of Condensed Matter Physics, vol. 15, no. 2, (**2025**), pp. 17-32.

[2] G. P. Malik and V. S. Varma, "On the Role of Chemical Potential in Determining the Temperature Dependent Critical Magnetic Fields and the Penetration Depth of Superconductors", World Journal of Condensed Matter Physics, vol. 14, no. 4, (**2024**), pp. 96-106.

[3] G. P. Malik and V.S. Varma, "A New Microscopic Approach to Deal with the Temperature and Applied Magnetic-Field Critical Current Densities of Superconductors", Journal of Superconductivity and Novel Magnetism, vol. 34, 1551 (**2021**), pp. 1551-1561.

[4] G. P. Malik, "Superconductivity: A New Approach Based on the Bethe-Salpeter Equation in the Mean-Field Approximation", in Series on Directions in Physica, vol. 21, World Scientific, Singapore, (**2016**).

[5] G. P. Malik, "On the equivalence of the binding energy of a Cooper pair and the BCS energy gap: A framework for dealing with composite superconductor", International Journal of Modern Physics B, vol. 24, no. 9, (**2010**), pp. 1159-1172.

[6] G. P. Malik and V.S. Varma, “On a New Number Equation Incorporating Both Temperature and Applied Magnetic Field and its Application to MgB”, Journal of Superconductivity and Novel Magnetism, vol. 33, (**2020**), pp. 3681-3685.

[7] T. Baumgartner et al., “Effects of Neutron Irradiation on Pinning Force Scaling in State-of-the-Art $Nb_3Sn$ Wires”, Superconductor Science and Technology, vol. 27, (**2014**), Article ID 015005.

[8 N. R. Werthamer, E. Helfand and P. C. Hohenberg, “Temperature and Purity Dependence of the Superconducting Critical Field, $H_{c2}$ III Electron Spin and Spin- Orbit Effects” Phys. Rev., vol. 147, (**1966**), pp. 295-302.

[9] C. K. Jones, J. K. Hulm and B. S Chandrasekhar, “Upper Critical Field of Solid Solution Alloys of the Transition Elements”, Rev. Mod. Phys., vol. 36, (**1964)**, pp. 74-76.

[10] L. P. Gorkov, “The Critical Supercooling Field in Superconductivity Theory”, Soviet Phys., JETP, vol. 37(10), no. 10, (**1960**), pp. 593-599.

[11] E. F. Talantsev, “Classifying Superconductivity in Compressed $H_3S$”, Modern Physics Lett. B, vol. 33, (**2019**), paper ID 1950195.

[12] V. S. Minkov et al., “Magnetic flux trapping in hydrogen-rich high-temperature superconductors”, Nature Physics, vol. 9, (**2023**), pp. 1293-1300.

[13] G. P. Malik and V.S. Varma, “Compressed H3S: Fits to the Empirical $H_{c2}(T)$ Data and a Discussion of the Meissner Effect”, World Journal of Condensed Matter Physics, vol. 13, no. 4, (**2023**), pp. 111-127.

[14] Charles P. Poole, Handbook of Superconductivity, 2ne edition, Academic Press, NY, (**2000**).

[15] C. Buzea and Y. Yamashita, “Review of the Superconducting Properties of $MgB_2$”, Superconductor, Science and Technology, vol. 14, R115 (**2001**), pp. R115-R146.

[16] I. S. Balbaa and F.D. Manchester, “Superconductivity of $PdH_x$ in relation to its phase diagram. I. Magnetic measurements”, Journal of Physics F: Metal Physics, vol. 13, no. 2, (**1983**), pp. 395-404.

[17] A. J. Leggett, PHYS598/2 Lecture 2: Non- Cuprate Exotics I: BKBO, $MgB_2$, Alkali Fullerides (courses.physics.illinois.edu) (**2018**)

[18] E.F. Talantsev, “Universal Fermi Velocity in Highly Compressed Hydride Superconductors”, Matter and Radiation at Extremes, 7, (**2022**), paper ID 058403.

[19] I. A. Kruglov et al., “Superconductivity of $LaH_{10}$ and $LaH_{16}$ polyhydrides”, Phys. Rev. B, vol. 101, no. 2 (**2020**), 024508 (pp. 8)

[20] B. J. Ramshaw et al., “Quasiparticle Mass Enhancement Approaching Optimal Doping in a High-$T_c$ Superconductor”, Science, 348.6232, 317, (**2015**), pp.317-320.

[21] G.P. Malik and V.S. Varma, “A Fermi Energy-Incorporated Framework for Dealing with the Temperature- and Magnetic Field-Dependent Critical Current Densities of Superconductors and Its Application to Bi-2212”, World Journal of Condensed Matter Physics, vol. 10, no. 2, (**2020**), pp. 53-70.

[22] L. Gunther and L. W. Gruenberg, “De Haas-Van Alphen Oscillations in the Critical Temperature of Type II Superconductors”, Solid State Communications, vol. 4, no. 7, (**1966**), 329-331.

[23] A. Aharoni, “Demagnetizing factors for rectangular ferromagnetic prisms”, J. Appl. Phys*,* vol. 83, no. 6, (**1998**), 3432-3434.

[24] G. S. Hosseinali, et al., “Critical currents in Tl-2212 and Tl-2223 thin films”, Physica C, vol. 268, (**1996**), pp. 307-316.

[25] V .S. Minkov, et al., “Magnetic field screening in hydrogen-rich high-temperature superconductors”, Nature Communications, 13, (**2022)**, 3194 (pp. 8)

[26] A. Audouard, et al., “Quantum oscillations and upper critical magnetic field of the iron-based superconductor FeSe”, Euro. Phys. Lett.”, vol. 109, (**2015**), 27003 (pp. 7)

[27] G. P. Malik, “An Overview of the Multi-Band and the Generalized BCS Equations-Based Approaches to Deal with Hetero-Structured Superconductors”, Open Journal of Microphysics, vol. 8, no. 2, (**2018**), pp. 7-13.

[28] M. Rasolt, Z. Tešanović, “Theoretical aspects of superconductivity in very high magnetic field”, Revs. Mod. Phys., vol. 64, (**1992**), pp. 709-754.

[29] G. P. Malik and V. S. Varma, “A Dynamical Approach to the Explanation of the Upper Critical Field

Data of Compressed $H_3S$", World Journal of Condensed Matter Physics, vol. 13, no. 3, (**2023**), pp. 79-89.